\documentclass{jfm}
\usepackage{graphicx}
\usepackage{epstopdf, epsfig}
\usepackage{amsmath}
\usepackage{comment}
\usepackage{subcaption}
\usepackage{natbib}
\usepackage{hyperref}
\hypersetup{
    colorlinks = true,
    urlcolor   = blue,
    citecolor  = black,
}
\usepackage{babel}
\usepackage{soul}
\setstcolor{red}

\usepackage{pgfplots}
\usepackage{pgfplotstable}
\usepgfplotslibrary{fillbetween}
\pgfplotsset{
  log x ticks with fixed point/.style={
      xticklabel={
        \pgfkeys{/pgf/fpu=true}
        \pgfmathparse{exp(\tick)}%
        \pgfmathprintnumber[fixed relative, precision=3]{\pgfmathresult}
        \pgfkeys{/pgf/fpu=false}
      }
  },
  log y ticks with fixed point/.style={
      yticklabel={
        \pgfkeys{/pgf/fpu=true}
        \pgfmathparse{exp(\tick)}%
        \pgfmathprintnumber[fixed relative, precision=3]{\pgfmathresult}
        \pgfkeys{/pgf/fpu=false}
      }
  }
}
\tikzset{decorated arrows/.style={
    postaction={
        decorate,
        decoration={
            markings,
            mark=at position 0.5 with {\arrow{stealth}}
            }
        },
    }
}
\pgfplotsset{select coords between index/.style 2 args={
    x filter/.code={
        \ifnum\coordindex<#1\def\pgfmathresult{}\fi
        \ifnum\coordindex>#2\def\pgfmathresult{}\fi
    }
}}
\pgfplotsset{compat=1.3}
\usepackage{tikz}
\usepackage{pgfplots}
\pgfplotsset{width=10cm,compat=1.9}
\usepgfplotslibrary{external}
\usetikzlibrary{external}
\usepackage{pgfplotstable}
\usepgfplotslibrary{fillbetween}
\usetikzlibrary{arrows.meta}
\usetikzlibrary{decorations.pathmorphing}
\usetikzlibrary{decorations.pathreplacing, arrows.meta}
\usetikzlibrary{decorations.markings}
\usetikzlibrary{patterns}
\usetikzlibrary{calc}
\pgfplotsset{compat=1.3}
\tikzset{external/system call={pdflatex \tikzexternalcheckshellescape -halt-on-error
-interaction=batchmode -jobname "\image" "\texsource"}}

\newcommand{\RomanNumeralCaps}[1]
 \linenumbers

\shorttitle{Free falling sphere motion in a deep pool}
\shortauthor{P. K. Billa, T. Josyula, C. Tropea and P. S. Mahapatra}

\title{Motion of a rigid sphere penetrating a deep pool}

\author{Prasanna Kumar Billa\aff{1},
    Tejaswi Josyula\aff{1,3}
 \and Cameron Tropea\aff{1,2}
  \corresp{\email{tropea@sla.tu-darmstadt.de}}, Pallab Sinha Mahapatra\aff{1}
  \corresp{\email{pallab@iitm.ac.in}}}
  
\affiliation{\aff{1} Multiscale Multiphysics Group (MMG), Department of Mechanical Engineering, Indian Institute of Technology Madras, 600036, India \aff{2} Institute for Fluid Mechanics and Aerodynamics, Technical University of Darmstadt, 64287, Germany  \aff{3} Current Address: FLOW, Department of Engineering Mechanics, KTH Royal Institute of Technology Stockholm, 11428, Sweden  }

\begin{document}
\maketitle

\begin{abstract}
In this  study, we experimentally examine the behavior of a free-falling rigid sphere penetrating a quiescent liquid pool. Observations of the sphere trajectory in time are made using two orthogonally placed high-speed cameras, yielding the velocity and acceleration vector through repeated differentiation of the time-resolved trajectories.

The novelty of this study is twofold. On the one hand, a methodology is introduced by which the instantaneous forces acting on the sphere can be derived by tracking the sphere trajectory. To do this, we work in a natural coordinate system aligned with the pathline of the sphere. In particular, the instantaneous lift and drag forces can be separately estimated. 

On the other hand, the results reveal that when decelerating, the sphere experiences a very high drag force compared with steady flow. This is attributed to an upstream shift of the mean boundary-layer separation. The sphere also experiences significant lift force fluctuations, attributed to unsteady and asymmetric wake fluctuations. The trajectories can be reduced to three stages, common in duration for all initial Reynolds numbers and density ratios when expressed in dimensionless time. In addition, the sphere velocity and deceleration magnitude for different initial parameters exhibit a high degree of uniformity when expressed in dimensionless form. This offers prediction capability of how far a sphere penetrates in time and the forces acting on it.

\end{abstract}


\section{\label{sec:introduction} Introduction}
Solid bodies impacting and penetrating into a quiescent liquid pool is a widely observed phenomenon with diverse practical applications in ship slamming \citep{zhao1993water, faltinsen1990sea}, boat hulls \citep{howison2002deep}, diving \citep{gregorio2023}, bullets \citep{Truscott2014WaterProjectiles}, underwater missiles \citep{may1975water}, air-to-sea anti-torpedo defense systems \citep{von1929impact,richardson1948impact,Truscott2009ASpheres},  and the transfer of solid objects to the liquid, like releasing oceanographic instruments into the sea \citep{Abraham2014ModelingEntry}. 
The penetration of solid spheres in a liquid pool, as investigated in the present study, represents a generic simplification of the above mentioned applications. In some instances, the water entry leads to the formation of a persistent air cavity in the wake of the sphere, depending on the boundary conditions and the wettability of the sphere \citep{worthington1883impact,May1951EffectCavity, Tan2016CavityLiquids,Aristoff2009WaterSpheres, Aristoff2010TheSpheres, Truscott2009ASpheres,truscott2012unsteady, Mansoor2014WaterFormation,Vakarelski2011DragLayers,mchale2009terminal, Mansoor2017Stable-streamlinedSpheres}.
What has not been fully elucidated is the necessary time or traversed distance before a flow around the sphere can be considered devoid of entry effects, even without an entrapped air cavity. The present study is restricted to impact conditions not resulting in such an air cavity.

The phenomenon of a rigid sphere traversing through a quiescent liquid at moderate Reynolds numbers
has been explored by \citet{kuwabara1983anomalous}. This study revealed that the spheres exhibited lateral motion away from a pure vertical trajectory. They attributed this to lateral/lift forces exerted on the sphere arising from asymmetric vortex shedding in the wake of the sphere. \citet{taneda1978visual} also studied this phenomenon using smoke flow visualization in a wind tunnel and concluded that a side/lift force acts on the sphere due to the asymmetric wake, something that had already been established by \citet{scoggins1967sphere}. These studies confirmed such a side force in the Reynolds number range $3.8 \times 10^5 < \mathrm{Re} < 10^6$, i.e., above the critical Reynolds number at which laminar-turbulent transition of the boundary layer occurs. 

The falling and rising of solid spheres in a quiescent liquid was investigated by \citet{Veldhuis2005MotionParticles} using the Schlieren technique for various solid-to-liquid density ratios ranging from 0.5 to 2.63 and various initial Re ranging from 200 to 4600. This study revealed that the path followed by a sphere  changes from a straight vertical line to a  deviation in a random direction. This was attributed to the formation of asymmetric vortices in the wake of the sphere.  \citet{horowitz2010effect} conducted an investigation into the behavior of spheres falling freely through a liquid with a relative density $\rho^\ast=$ $\rho_\mathrm{s}/\rho_\mathrm{l} > 1$, where $\rho_\mathrm{s}$ is the density of the sphere and $\rho_l$ is the density of the liquid. The study covered a range of Reynolds numbers (100 $<$ Re $<$ 15000) and found that the vortex shedding and wake patterns significantly influence the motion of the spheres. Subsequently, \citet{horowitz2010effect} undertook a comprehensive investigation of vortex formation in the wake of spheres and their dynamics within the liquid, illustrating the wakes and paths of solid spheres using regime maps that delineate distinct motion patterns including vertical, oblique, intermittent oblique, and zigzag trajectories. \citet{Ern2011Wake-inducedFluids} explored the kinematics and dynamics of spheres moving along irregular paths. The study revealed a close connection between the path instabilities of bluff bodies submerged in viscous liquids and the initiation of instability in the fixed-body wake. The research determined that vortex shedding in the wake plays a crucial role in inducing path instabilities in spheres, causing them to follow irregular trajectories. \citet{truscott2012unsteady} conducted a comprehensive study delving into the unsteady forces exerted on spheres of different densities as they impacted and penetrated a quiescent liquid pool. They successfully developed a technique to estimate hydrodynamic forces by utilizing both position data and acceleration, which were derived from the trajectory data by fitting of spline curves. The computed drag force was confirmed using PIV measurements in the wake to estimate circulation; hence, change of impulse force in time.

What is not consistently reported in the literature is whether spheres at higher Reynolds numbers exhibit spiraling motion when descending through the liquid. Both \citet{shafrir1965horizontal} and \citet{christiansen1965effect} observe corkscrew or spiraling trajectories of the spheres, whereas \citet{kuwabara1983anomalous} observe these very seldom. This is insofar an interesting phenomenon since such trajectories infer a sustained lateral/lift force on the sphere, otherwise, the trajectory would transition to a pure vertical settling motion.

While there is extensive literature on experimental studies, there have also been numerous studies devoted to theoretical and numerical aspects of this problem. By employing the Verlet algorithm, \citet{Valladares2003SimulationFluid} numerically investigated the motion of a solid sphere traveling through a viscous fluid, and the terminal settling velocities of the spheres were computed for varied viscosity of the fluid and density of the sphere. A complete analytical solution of the sphere falling through the liquid is given by \citet{Guo2011MotionFluids} for various Reynolds numbers. They have considered a rectilinear fall of a sphere in a quiescent fluid. The Basset–Boussinesq–Oseen (BBO) equation was solved for the acceleration of the sphere inside the viscous liquid.

Despite the numerous previous studies of a sphere moving through a liquid pool, the magnitude of the lateral/lift forces acting on the sphere to divert its trajectory from pure vertical motion have not yet been quantitatively reported. While there is general agreement that unsteady vortex shedding in the wake leads to these asymmetric forces, neither their frequency of occurrence nor their sustainability have been quantitatively addressed. Furthermore, there exists general agreement on the fact that a decelerating sphere can exhibit much higher drag forces than in steady flow, although  this phenomenon has also not been widely quantified or explained.

In the present study, we address these knowledge gaps using a novel approach to measuring the time-resolved forces acting on the sphere.
Two synchronised cameras are placed orthogonal to each other, capturing the three-dimensional sphere trajectory in time. This allows the instantaneous acceleration vector to be computed; hence, the acting force vector.  Working in a natural coordinate system, a force balance using the BBO equation yields a quantitative estimate of the instantaneous drag and lift forces (as dimensionless coefficients) acting on the sphere during its penetration trajectory. 

The insight gained using this methodology on the one hand reveals a remarkably uniform collapse of the motion kinematics over all investigated impact parameters when expressed in dimensionless form. Moreover, the time-resolved lift and drag coefficients obtained under strong decelerating conditions, suggest a certain specific behaviour of the boundary-layer separation under these conditions.

\section{\label{sec:experimental details}Experimental setup and methodology} 
\subsection{\label{sec:experimental setup}Experimental setup}
The experimental setup consists of a clear, translucent acrylic container and high-speed monochrome cameras. The cross-section of the container is large,  200 $\times$ 200 $\mathrm{mm}^2$, 20 times larger than the diameter of the largest sphere utilized in the present study.  The depth of the container is 400 mm. By creating a suction pressure at the end of a needle tip, the spheres are firmly held and are released by interrupting this suction pressure. The free falling sphere impacts the liquid with no rotation, as confirmed by the images captured prior to sphere impact. All the experiments are performed in a closed room with an ambient temperature of 25$^\circ$C. 

\begin{figure}
\centering
\includegraphics[width=\textwidth]{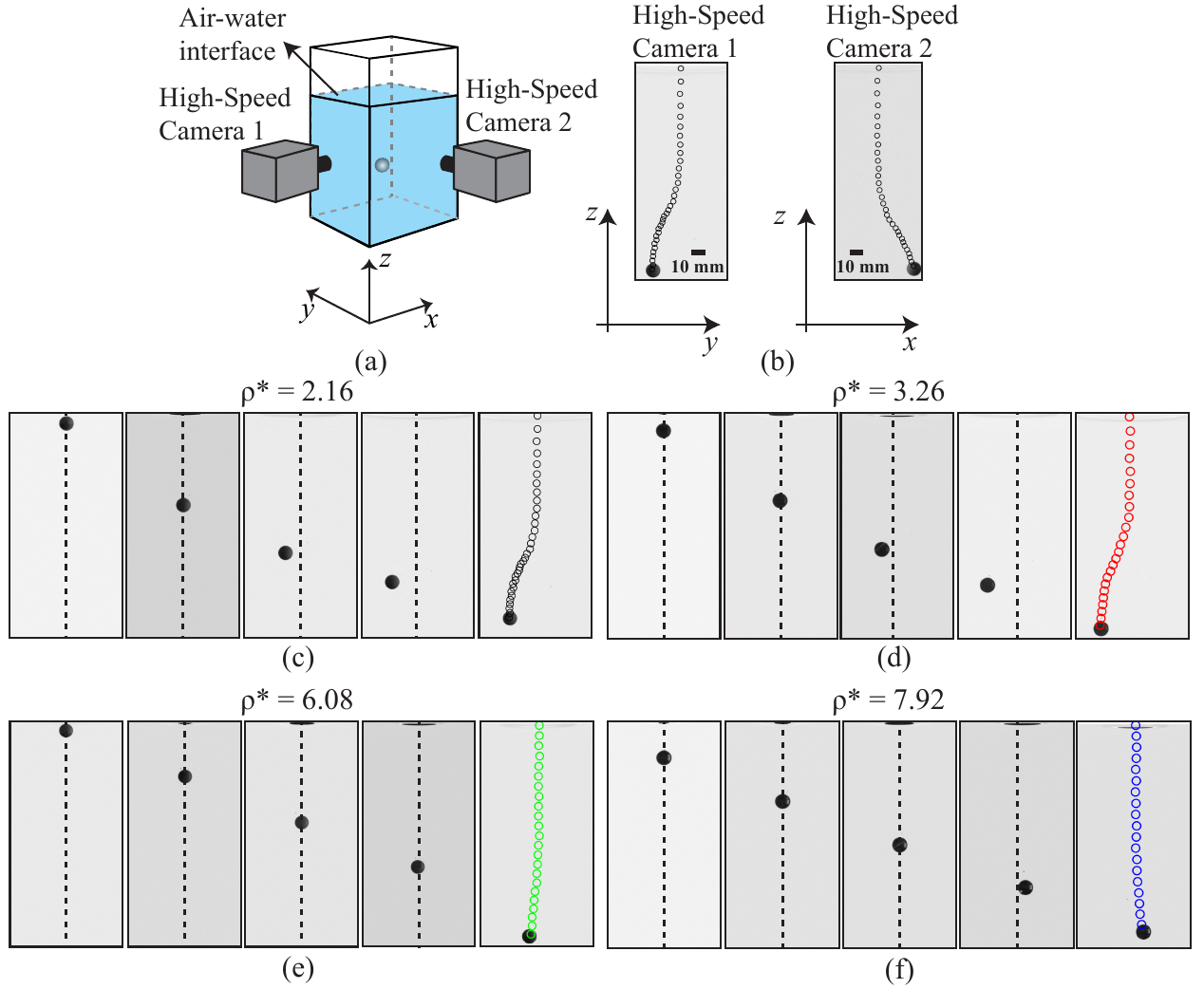}
\caption{(color online) Two orthogonally and synchronized cameras  record the motion of a rigid sphere after impacting the air-water interface. (a) illustrates schematically the experimental arrangement and the coordinate system used for the imaging data;  (b) a sample trajectory of a sphere with a density ratio ($\rho^*$) of 2.16 and diameter 10~mm is viewed by the two cameras in the $z$-$y$ and $z$-$x$ planes. Image sequences of 10~mm spheres with an initial Reynolds numbers of 15700 are shown for the density ratios (c) $\rho^* = 2.16$, (d) $\rho^* = 3.26$, (e) $\rho^* = 6.08$, and (f) $\rho^* = 7.92$. The time step between consecutive images is 60 ms for (c), 40 ms for (d), 25 ms for (e), and 20 ms for (f). The black dashed line represents a pure vertical trajectory. The circular data points in the last images indicate the position of the sphere at equal time intervals.}
\label{fig;two cameras}
\end{figure} 

Experiments were conducted using two cameras (Phantom VEO E-340-L with  Nikon microlenses of 28 mm, spatial resolution of 192 $\mu$m/pixel and  24-85 mm, spatial resolution of 199 $\mu$m/pixel) placed orthogonal to one another. Backlighting consisted of a 30~W monochromatic light source and a diffuser.
The two cameras were synchronised and recorded the sphere motion at 1000 fps, with a resolution of 1152 $\times$ 1100 pixels (100~mm $\times$ 160~mm in the object plane) and an exposure time of 400 $\mu s$. The exposure time leads to a maximum relative motion blur of 28\% of the sphere diameter for the smallest sphere (4~mm) with the highest initial velocity (2.8~m/s). However, the motion blur for the 10~mm sphere is only 11\% upon impact, and for all spheres, this motion blur  decreases rapidly after impact since the velocity immediately goes through a strong deceleration phase. Furthermore, the edge detection routine remains the same at all time steps, so the error through motion blur for the relative motion is neglected.  The data is collected while the sphere descends until it traverses out of the field of view of either of the cameras. A pictorial view of the experimental setup is shown in Fig. \ref{fig;two cameras}(a). Figure \ref{fig;two cameras}(b) shows a sample trajectory of a 10~mm sphere with density ratio of $\rho^*=$ 2.16, as observed with the two cameras. 

To further illustrate the raw data with which we are working,  we present in Fig.~\ref{fig;two cameras}(c-f), showing example trajectory traces from a single camera for the 10~mm diameter spheres, all with the same initial Reynolds number ($Re_i$ defined in Table~\ref{tab:parameters}), but for four different density ratios ($\rho^*$). It is apparent from these visualizations that the lighter spheres (c, d) exhibit higher lateral displacements away from the vertical, whereas the heavier spheres (e, f) have a more ballistic-like trajectory, as expected. The associated lift forces causing these lateral displacements will be quantitatively derived below. 

Table~\ref{tab:parameters} provides the parameters and dimensionless quantities pertinent to this study. The diameter ($D$) of the spheres is measured using a Vernier caliper and their mass ($m$) is determined using an electronic weight balance (Ohaus), yielding their density   $\rho_\mathrm{s}$. The impact velocity ($v_\mathrm{i} \approx \sqrt{ 2gh }$) is derived from the initial release height ($h$) of the sphere, measured from its center to the air-water interface. The Reynolds number ($Re_i$=$v_\mathrm{i}D/\nu$) upon impact ranges from 6300 to 31500. Given that the critical Reynolds number for a sphere in steady flow lies well above 10$^5$, we assume that the boundary layers on all spheres throughout all phases of the trajectory remain laminar.

For these definitions, $\nu$ is the kinematic viscosity (0.89 $ \mathrm{mm^2/s}$) of the fluid, and $g$ is the gravitational acceleration.
Throughout the following discussion, length scales are rendered dimensionless using the sphere diameter $D$, velocities with the impact velocity $v_\mathrm{i}$, and time scales using $D/v_\mathrm{i}$. Dimensionless quantities are designated with the superscript `$*$', and unit vectors are written in boldface font.

This range of Reynolds number was chosen, because sphere impact at lower initial Reynolds numbers showed no significant difference in trajectory behaviour than those conducted, and higher Reynolds number impacts led to an entry air cavity being formed behind the penetrating sphere. The entry phenomenon was outside the scope of the present study, but has been investigated at length in connection with the impact and penetration of superhydrophic spheres \citep{Speirs2019WaterAngles}. The density ratio was not chosen beyond 7.92, since at this value, no significant change was observed from the value 6.08. At these high density ratios the sphere exhibits a nearly vertical trajectory, which is seen in Fig.~\ref{fig;two cameras}(e, f).

\subsection{Methodology}
\begin{table}
  \begin{center}
   \caption{\label{tab:parameters}%
    Definitions and range of parameters 
    }
\def~{\hphantom{0}}
  \begin{tabular}{c c c c}
  \hline 
     \textrm{Parameter}   &  \textrm{Symbol}  &   \textrm{Definition}   &  \textrm{Range of values}  \\[3pt]
     \hline 
                 Sphere diameter & $D$ & -  & 4, 6, 10~mm  \\
                 Density ratio &$\rho^*$ & $\rho_\mathrm{s}/\rho_\mathrm{l}$  &  2.16, 3.26, 6.08, 7.92  \\
                Impact velocity & $v_i$ & $ \sqrt{2gh} $  & 1.40, 1.98,  2.80 m/s \\
                 initial Reynolds number &$Re_{i}$ & $v_\mathrm{i}D/\nu$  & 6300 - 31500  \\          
   \hline
  \end{tabular}
  \end{center}
\end{table}

\begin{figure}
    \centering
    \includegraphics[width=0.7\textwidth]{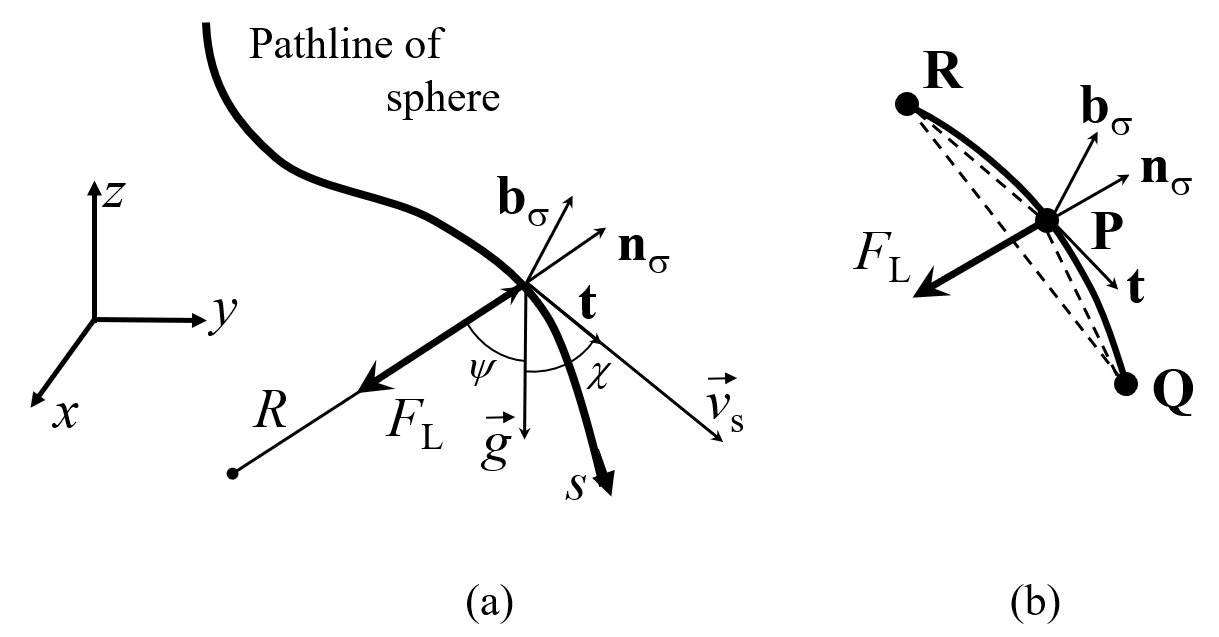}
    \caption{(a) Definition of the natural coordinate system based on the sphere pathline.
    (b) Definition of the plane of curvature and the total lift magnitude $F_\mathrm{L}$.  }
    \label{fig:pathline}
\end{figure}

For the presentation of results and the subsequent analysis invoking a force model, it is convenient to work in a natural coordinate system, i.e., in a coordinate system in which the unit vectors of the accompanying triad of the pathline are used as basis vectors. This coordinate system is pictured in Fig.~\ref{fig:pathline}(a). The unit vector tangential to the pathline is given as 
\begin{equation}
    \label{eq:unit vector pathline}
    \mathbf{t}=\frac{\Vec{v}_\mathrm{s}}{| \Vec{v}_\mathrm{s} |}
\end{equation}
where $\Vec{v}_\mathrm{s}$ is the velocity vector of the sphere along the pathline $s$. $\sigma$ is the coordinate in the direction $\mathbf{t}$, $n$ is the coordinate in the direction of the principal normal vector $\mathbf{n}_\sigma = R\mathrm{d}\mathbf{t}/\mathrm{d}\sigma$, and $b$ the coordinate in the direction of the binormal unit vector $\mathbf{b}_\sigma=\mathbf{t} \times \mathbf{n}_\sigma$. $R$ is the radius of curvature of the pathline in the plane spanned by the normal vectors $\mathbf{t}$ and $\mathbf{n}_\sigma$. The velocity vector  $\vec{v}_\mathrm{s}$ is understood to also represent the slip velocity in the equation of motion, since the pool is quiescent upon sphere impact. 

\subsection{Image Processing}
\label{subsection:image processing}
The video images are processed using ImageJ software to subtract the background and to create a binarized image of the sphere inside the liquid. Subsequently, an in-house Matlab\textsuperscript{\tiny\textregistered} code is employed to determine the position of the sphere.  The detection of the bottom most point of the sphere is accomplished by utilizing an edge detection technique. The air-water interface is established as the reference point for spatial coordinates ($z=0$).
Similarly, the time instant a sphere makes first contact with the air-water surface is established as the reference point for time ($t=0$). This time is taken as the first frame in which contact with the liquid has been made.  

  The accumulated dimensionless path length that the sphere covers is denoted $s^*=s/D$, whereby $s=0$ at the air-water interface.  The dimensionless lateral displacement at each time step is expressed as $r^*=\sqrt{x^2+y^2}/D$. \textcolor{black}{The displacement data is smoothed using an in-house MATLAB function  'loess', a method that involves using linear regression in a locally weighted scatter plot. Subsequently, velocity magnitude $|\Vec{v}_\mathrm{s} | $ is computed by differentiating the smoothed displacement data after first fitting the data with a quintic spline function.} The dimensionless velocity magnitude is given as $v_\mathrm{s}^*=| \Vec{v}_\mathrm{s}|/v_\mathrm{i}$.  Acceleration data was obtained using a differentiation of the smoothed velocity data. Time is made dimensionless using the diameter of the sphere and the impact velocity, i.e., $t^*=t v_\mathrm{i}/D$. 

  Given the $x$, $y$ and $z$ coordinates of the sphere as a function of time, the curvature of the pathline can be computed as 
  \begin{equation}
     \label{eq:curvature}
     \kappa = \frac{1}{R} = \frac{\sqrt{(z''y'-y''z')^2+(x''z'-z''x')^2+(y''x'-x''y')^2}}{(x'^2+y'^2+z'^2 )^{3/2}}
 \end{equation}
 where the primes indicate first and second differentiation with respect to time. These differentials are computed numerically using second-order central differences.  The angular frequency along the curved pathline is then given as  $\omega = | \Vec{v}_\mathrm{s}|/R$, which is a necessary quantity in computing the centrifugal force acting on the sphere, which will be related to the lift force, examined in the following section.

 \subsection{Force model}
\label{subsec:force model}
In formulating the equation of motion for the sphere, the Boussinesq-Basset-Oseen (BBO) equation  \citep{zhu1998multiphase}  is used, which includes body forces ($F_\mathrm{G}$ - weight, $F_\mathrm{B}$ - buoyancy), apparent forces ($F_\mathrm{H}$ - Basset or history term, $F_\mathrm{A}$ - added mass), hydrodynamic forces ($F_\mathrm{D}$ -  viscous and pressure forces combined as drag) and inertial forces ($F_\mathrm{I}$). We will neglect the Saffman lift force \citep{saffman1965lift}, applicable only in sheared flow, and any rotational-lift force \citep{rubinow1961transverse}, applicable only with rotation/spin of the sphere.

The scalar momentum equation expressed along the direction of motion/pathline can be written as \citep{crowe2011multiphase}
\begin{eqnarray}
      \underbrace{\frac{1}{6} \rho_\mathrm{s} \pi D^3 \frac{\mathrm{d}v_\mathrm{s}}{\mathrm{d}t}}_\text{Inertial force}=       
      \underbrace{\frac{1}{6} \pi D^3 (\rho_\mathrm{s} - \rho_\mathrm{l}) g \cos{\chi}}_\text{Body forces}        - \underbrace{\frac{1}{8} C_\mathrm{D} \rho_\mathrm{l} \pi D^2 v_{s} ^2}_\text{Drag force} 
      - \underbrace{\frac{1}{6} C_\mathrm{A} \rho_l \pi D^3 \frac{\mathrm{d}v_\mathrm{s}}{\mathrm{d}t}}_\text{Added mass} \nonumber \\ 
      - \underbrace{\frac{3}{2} D ^2 \sqrt{\pi \mu \rho_\mathrm{l}} \int_{0}^{t} \frac{1}{\sqrt{t-\zeta}} \frac{\mathrm{d}v_\mathrm{s}}{\mathrm{d}\zeta}\mathrm{d}\zeta }_\text{Boussinesq-Basset term} 
      \label{eq:Force evaluation} 
\end{eqnarray}

\noindent where   $C_\mathrm{D}$ is the coefficient of  drag,
and $C_\mathrm{A}$ is the coefficient of the added mass force.  Note that no hydrodynamic lift force  has been included in Eq.~(\ref{eq:Force evaluation}). This force will be introduced and discussed below.

The added mass coefficient $C_\mathrm{A}$ is usually taken as 0.5 \citep{Guo2011MotionFluids} and expresses the kinetic energy imparted into the surrounding fluid through acceleration/deceleration of the sphere. The Boussinesq-Basset term captures the viscous force change due to boundary-layer development on an accelerating or decelerating submerged body. These viscous forces are not expected to be significant relative to the inertial and pressure forces involved over large portions of the trajectory; hence, the Boussinesq-Basset term will be neglected in the following analysis \citep{Nouri2014UnsteadyMethods}. This is not to say that the transient boundary layer development does not play a central role in determining the trajectory and speed of the sphere, but it is expected to be more through the separation and wake behaviour due to the state of the boundary layer; hence, through the resulting pressure distribution around the sphere. These forces would make themselves apparent in the above equation as variations in the drag force.

The initial Reynolds numbers encountered in this study all lie below approx.  $3 \times 10^4$ and throughout most of the sphere trajectory the values are much lower. For very similar Reynolds numbers \citet{truscott2012unsteady} used a drag coefficient of 0.5. They viewed the unsteady added mass as part of the pressure force acting on the sphere, i.e., in the net hydrodynamic force.
 In the present study, the experimental data allows all of the remaining terms in Eq.~(\ref{eq:Force evaluation})  to be evaluated; thus, the value of $C_\mathrm{D}$ can and will be computed at each time step. 

 The terminal velocity of a sphere can be determined by establishing an equilibrium in which the net force acting on the sphere is reduced to zero. The expression for the dimensionless terminal velocity ($v_\mathrm{t}^*$) can be obtained by equating the buoyancy force with the  drag force plus the weight:
\begin{equation}
    v_\mathrm{t}^* = \frac{1}{v_\mathrm{i}}\sqrt{ \frac{4 (\rho^* - 1)gD }{3C_\mathrm{D}} }
    \label{eq; terminal velocity}
\end{equation}
In this equation, a drag coefficient must be prescribed, and a value of $C_\mathrm{D}=0.5$ has been used, assuming at this stage no significant acceleration or deceleration.

If the pathline has  non-zero curvature ($\kappa$), this implies a  force ($\overrightarrow{F}_\mathrm{L}$) acting in the $-\mathbf{n}_\sigma$ direction, i.e., perpendicular to $\mathbf{b}_\sigma$, as depicted in Fig.~\ref{fig:pathline}(b). 
Since lift force is defined as the force acting perpendicular to the direction of motion, the total lift force magnitude, $||\overrightarrow{F}_\mathrm{L}||=F_\mathrm{L}$,  can be computed as the sum of the buoyancy ($F_\mathrm{B}$), gravity ($F_\mathrm{G}$), and the hydrodynamic (lift)force ($F_\mathrm{HL}$) acting in the $-\mathbf{n}_\sigma$ direction. Thus, to compute the hydrodynamic lift force from the sphere trajectory, the direction of the unit normal $\mathbf{n}_\sigma$ must be determined. Recognizing that $\mathbf{n}_\sigma=\mathbf{b}_\sigma \times \mathbf{t}$, it is necessary to first compute the unit normal $\mathbf{b}_\sigma$ from the trajectory data. This is done using three points along the trajectory, pictured in Fig.~\ref{fig:pathline}(b) as points \textbf{PQR}, whereby point \textbf{P} is the position at which the local force is to be estimated. \textcolor{black}{Point \textbf{R} represents the position just before point \textbf{P}, and point \textbf{Q} is the  position following point \textbf{P} along the trajectory.  \textcolor{black}{The vector $ \overrightarrow{\mathbf{PQ}} $  is derived by computing the forward difference between the coordinates of point \textbf{P} and point \textbf{Q}. Similarly, the vector $ \overrightarrow{\mathbf{PR}} $ is determined from the points \textbf{P} and \textbf{R}. This procedure is applied at every point along the trajectory of the sphere from the air-water interface.} }

Knowing the $x,y,z$ coordinates of the points \textbf{P}, \textbf{Q} and \textbf{R}, the unit normal to the subscribed triangle is given by  $\mathbf{b}_\sigma = \overrightarrow{\mathbf{PR}} \times \overrightarrow{\mathbf{PQ}}/||\overrightarrow{\mathbf{PR}}\times \overrightarrow{\mathbf{PQ}}||$. The trajectory unit normal $\mathbf{t}$, can be computed using forward differencing around point \textbf{P} allowing  $-\mathbf{n}_\sigma=-\mathbf{b}_\sigma \times \mathbf{t}$ to be computed.
The angle $\psi$ is the angle between $-\mathbf{n}_\sigma$ and the direction of gravity (see Fig.~\ref{fig:pathline}(a)). Only the components of buoyancy and gravity along the direction $-\mathbf{n}_\sigma$ can contribute to the total magnitude of the lift force $F_\mathrm{L}$, which itself must be equal to the centripetal force; hence, 
$(F_\mathrm{G}-F_\mathrm{B})\cos{\psi}+F_\mathrm{HL}=F_\mathrm{L}=m\omega^2R$, and with the angular frequency  given by $\omega = | \Vec{v}_\mathrm{s} |/R$
\begin{equation}
\label{eq:centripetal force}
    (F_\mathrm{G}-F_\mathrm{B})\cos{\psi}+F_\mathrm{HL}=\frac{m | \Vec{v}_\mathrm{s} |^2}{R}
\end{equation}
Note that this equation is a scalar equation, since through the $\cos\psi$ factor, the gravitational and buoyancy forces have been projected onto the -$\mathbf{n}_\sigma$ vector. Moreover, although the total lift force defined in this manner will always be positive and directed towards the origin of the local curvature radius, as shown in Fig.~\ref{fig:pathline}(b), the hydrodynamic lift force could become negative. A negative hydrodynamic lift force would act to reduce the local curvature of the sphere trajectory, i.e., straighten the trajectory. If now the buoyancy and gravity forces are known, then the hydrodynamic portion of the lift force $F_\mathrm{HL}$ can be computed and plotted as a function of time or displacement of the sphere. The hydrodynamic lift force is made dimensionless in the form of a lift coefficient, $C_\mathrm{L}$, 
 \begin{equation}
     \label{eq:lift coefficient}
     C_\mathrm{L}=\frac{F_\mathrm{HL}}{\frac{1}{2}\rho_\mathrm{l}| \Vec{v}_\mathrm{s} |^2 A}
 \end{equation}
 where $A$ is the projected area in the direction of motion, in this case $\pi D^2/4$. In some of the results presented below, an abrupt jump in lift coefficient was observed. An explanation of this result is given in the appendix.

Before presenting the results of the experiments, the origin of a lift force arising from an asymmetric wake and the interrelation between drag and lift from such wakes will be phenomenologically discussed. 
The initial Reynolds number lies in the approximate range $6300 < Re_i < 31500$; however, the sphere decelerates during its trajectory to values of approximately 10\% of the impact velocity, thus, the total encountered Reynolds number range is approximately $630 < Re_i < 31500$. This range lies in the Newton regime of drag coefficient, in which $C_\mathrm{D}$ takes an almost constant value over all Reynolds numbers and which is significantly below transitional Reynolds numbers for even roughened spheres. On the other hand, the data on which this statement is based comes from experiments in which the flow is steady, i.e., no acceleration or deceleration. Thus, accepted drag coefficients for steady flow may not necessarily be applicable over the entire sphere trajectories of the present experiments.

Although the flow and wake may be statistically symmetric around the sphere in this Reynolds number range, the instantaneous wake can be highly asymmetric. Thus, even though the time averaged lift coefficients in this Reynolds number range may be zero, instantaneous fluctuations of lift can occur and have been demonstrated through numerical simulations \citep{yun2006vortical,constantinescu2004numerical}. The transition of the boundary layer to a turbulent state after separation from the sphere can be irregular, causing it to temporarily reattach to the surface of the sphere and separate further downstream \citep{hadvzic2002experimental}.   Both \citet{taneda1978visual} and \citet{hadvzic2002experimental} observe a progressive wave motion around the sphere for $10^4 < \mathrm{Re}<3.8 \times 10^5$ by which the separation points rotate around the sphere randomly. \citet{achenbach1974vortex} also observed this for a very similar Reynolds number range. Such an irregular boundary-layer separation will also affect the drag, since the drag arises from the integration of the pressure around the sphere and is thus highly correlated with the location of flow separation:  a fluctuating separation will yield a fluctuating drag force.  However, an asymmetric separation and vortex shedding will not only influence the drag, but the asymmetry will also mean that the resultant force will no longer be aligned with the flow direction; this force will therefore, have both drag and lift components. For a free moving body this  results in a change of trajectory and a new orientation of the drag and lift forces in a lab-fixed coordinate system.

In summary, our force model is cast in a natural coordinate system and all changes in the motion speed of the sphere are attributed to a change in drag plus the body forces along the direction of motion. All direction changes are attributed to a hydrodynamic lift force plus the body forces acting in the direction of local pathline curvature. A positive lift force increases the pathline curvature, a negative lift force decreases the pathline curvature.

\section{ Measurement Results}
\label{sec:measurement results}
\subsection{Kinematics of sphere motion}
\label{subsection:kinematics}

\begin{figure}
    \begin{subfigure}{0.45\textwidth}
    \includegraphics[width=0.9\textwidth]{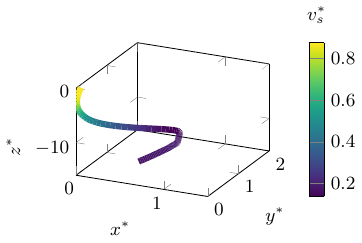}
    \caption{}
  \end{subfigure}
  \begin{subfigure}{0.45\textwidth}
    \includegraphics[width=0.9\textwidth]{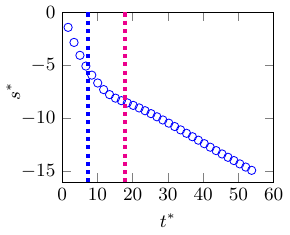}
    \caption{}
  \end{subfigure}
  
  \begin{subfigure}{0.45\textwidth}
    \includegraphics[width=0.9\textwidth]{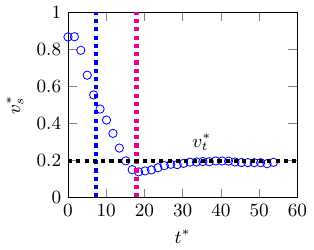}
    \caption{}
  \end{subfigure}
  \begin{subfigure}{0.45\textwidth}
    \includegraphics[width=0.9\textwidth]{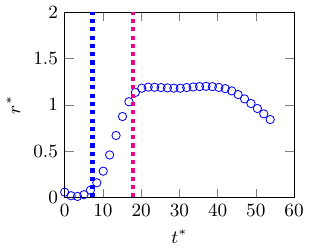}
    \caption{}
  \end{subfigure}
    \caption{(color online) Motion kinematics of a 10~mm sphere ($\rho^*=2.16$) entering the pool with an initial Reynolds number of 31500. (a) visualization of three-dimensional trajectory;  (b) dimensionless pathline distance over dimensionless time;  (c) instantaneous dimensionless velocity over dimensionless time; (d) dimensionless lateral deviation over dimensionless time.  The horizontal dashed line in this graph represents the terminal velocity computed according to Eq.~(\ref{eq; terminal velocity}). The dimensionless velocity ($v_s^*$)  of the sphere is represented by the colour bar in (a). Data points are spaced equally in time at 6~ms.}
    \label{fig:sample kinematics}
\end{figure}

Before discussing the results of the parametric variations conducted in this study, a sample set of data will be examined to illustrate the kinematics of the sphere motion, as shown in Fig.~\ref{fig:sample kinematics}. This figure shows a three-dimensional rendition of the trajectory (see Fig.~\ref{fig:sample kinematics}(a)), dimensionless pathline distance ($s^*$) (Fig.~\ref{fig:sample kinematics}(b)),  dimensionless instantaneous velocity magnitude ($v_s^*$) (Fig.~\ref{fig:sample kinematics}(c)), and the dimensionless lateral deviation $r^*$ (Fig.~\ref{fig:sample kinematics}(d)) for a 10~mm sphere ($\rho^* = 2.16$) impacting the pool with a Reynolds number of 31500. The dimensionless velocity magnitude ($v_s^*$) of the sphere in a three-dimensional rendition is depicted using a color bar in Fig.~\ref{fig:sample kinematics}(a). Two dashed lines have been added to graphs (b)-(d), the first denoting the dimensionless time at which the sphere first deviates more than 10\% of its diameter from the vertical trajectory ($r^*>0.1$), and the second denoting the time at which the sphere attains a minimum dimensionless velocity ($v_s^*$). These timelines divide the total time span into three phases, designated here as \\

\begin{itemize}
    \item \textbf{Submersion phase}, exhibiting a nearly vertical trajectory and extending up to the time at which the lateral displacement remains below  10\% of the diameter of the sphere.
    \item \textbf{Deceleration phase}, during which the sphere velocity continues to decrease to a minimum value. 
    \item \textbf{Settling phase}, the remaining time during which the sphere velocity tends towards its nominal terminal velocity in the vertical direction\\
\end{itemize}

\begin{figure}
    \begin{subfigure}[b]{0.3\textwidth}
    \includegraphics[trim=-0.24cm 0 0 0cm, width=0.99\textwidth]{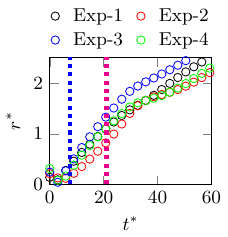}
    \caption{}
  \end{subfigure}
  \begin{subfigure}[b]{0.3\textwidth}
  \centering
   \hspace{5cm} \includegraphics[width=0.8\textwidth]{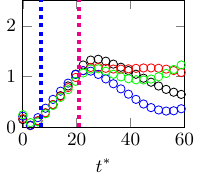}
    \caption{}
    \end{subfigure}
    \begin{subfigure}[b]{0.3\textwidth}
    \centering
    \includegraphics[width=0.8\textwidth]{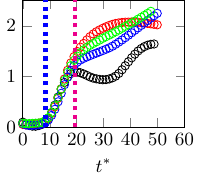}
    \caption{}
    \end{subfigure}
	 \vspace{1em}    
    
    \begin{subfigure}[b]{0.3\textwidth}
    \centering
    \includegraphics[trim=0.22cm 0 0 0cm, width=\textwidth]{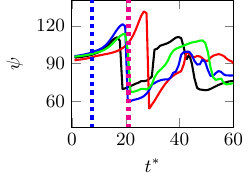}
    \caption{}
    \end{subfigure}
    \begin{subfigure}[b]{0.32\textwidth}
    \centering
    \includegraphics[trim=0.5cm 0 0 0cm, width=0.705\textwidth]{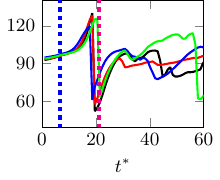}
    \caption{}
    \end{subfigure}
    \begin{subfigure}[b]{0.32\textwidth}
    \centering
    \includegraphics[trim=0.96cm 0 0 0cm, width=0.62\textwidth]{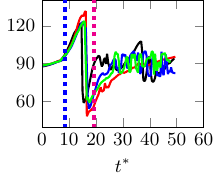}
    \caption{}
    \end{subfigure}
     \vspace{1em}
    
    \begin{subfigure}[b]{0.32\textwidth}
    \includegraphics[width=0.95\textwidth]{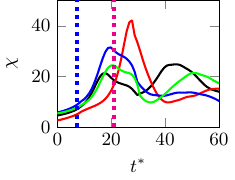}
    \caption{}
    \end{subfigure}
    \begin{subfigure}[b]{0.3\textwidth}
    \includegraphics[width=0.85\textwidth]{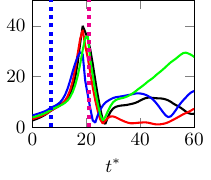}
    \caption{}
    \end{subfigure}
    \begin{subfigure}[b]{0.3\textwidth}
    \includegraphics[width=0.87\textwidth]{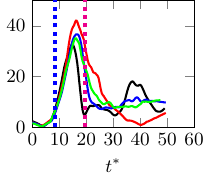}
    \caption{}
    \end{subfigure}
    \caption{(color online) Examples of measured dimensionless lateral deviations ($r^*$), angle $\psi$ and angle $\chi$  when repeating an experiment: (a,d,g) 4~mm sphere at $Re_i = 8900$; (b,e,h) 6~mm sphere at $Re_i = 18900$; (c,f,i) 10~mm sphere at $Re_i = 22300$. All spheres have $\rho^*=2.16$. The time step between the two successive points in (a), (b), and (c) is 6~ms. For clarity, angle $\psi$ is plotted as lines in (d)-(f), and angle $\chi$ is plotted as lines in (g)-(i).} 
    \label{fig:sample repetetion}
\end{figure}

Notably, the end of the deceleration phase (minimum velocity) coincides with the time at which the lateral deviation of the pathline ($r^*$) begins to exhibit strong variations among repetitions of the same experiment. This behavior is illustrated in Fig.~\ref{fig:sample repetetion}, in which typical lateral deviation curves are shown for repeated experiments using the 4~mm, 6~mm, and 10~mm diameter spheres with $\rho^*=2.16$.  
Any changes in $r^*$ must necessarily be associated with a lift force, as discussed above.

In Fig.~\ref{fig:sample repetetion} the variation of the angles $\psi$ (d-f) and $\chi$ (g-i) are also shown over dimensionless time. These results will be further discussed in section~\ref{subsection:dynamics}; however, their physical interpretation is briefly described here. The angle $\chi$ is simply the angle between the instantaneous pathline and the vertical. It starts with the value zero and ends at zero if the sphere has reached its terminal velocity downward. The angle $\psi$ is the angle between the plane of pathline curvature and gravity; hence, this angle expresses to what extent the body forces contribute to instantaneous lift. This angle must start at 90$^\circ$ and would also end at 90$^\circ$ if the sphere was moving vertically. 

\begin{figure}
    \begin{subfigure}[b]{0.33\textwidth}
   \includegraphics[width=\textwidth]{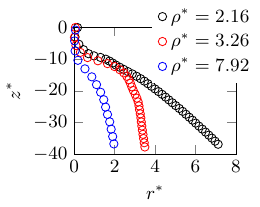}
 \caption{}
   \end{subfigure}
 \begin{subfigure}[b]{0.32\textwidth}
 \includegraphics[width=0.9\textwidth]{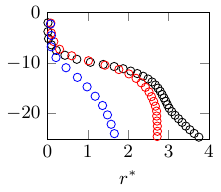}
         \caption{}
    \end{subfigure}
    \begin{subfigure}[b]{0.32\textwidth}
 \includegraphics[width=0.85\textwidth]{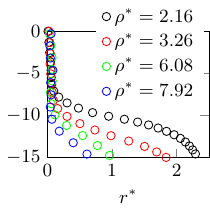}
         \caption{}
    \end{subfigure}
    \caption{(color online) Sample dimensionless lateral deviation ($r^*$) against corresponding dimensionless depth ($z^*$) for spheres of various density ratios.    (a) 4~mm sphere at $Re_i = 6300$; (b) 6~mm sphere at $Re_i = 9400$; (c) 10~mm sphere at $Re_i =15700$. The time step between trajectory points is 10~ms. }
    \label{fig:sample density deviations}
\end{figure}

The interpretation of $r^*$ is further illustrated and discussed in the dimensionless trajectory plots shown Fig.~\ref{fig:sample density deviations}. Here the dimensionless lateral deviation $r^*$ is plotted against the dimensionless depth $z^*$ for several density ratios $\rho^*$ and for the three sphere diameters. Note that the axes in these graphs vary since, in dimensionless terms, the observation volume of the acrylic container depends on the sphere diameter. It is apparent that not all of the spheres reach this final state within the available observation volume of the acrylic container. On the other hand, this is not considered a limitation, since, as will be shown in the next section, the fluctuating lift force has decreased significantly before the terminal velocity is reached.

\begin{figure}
 \includegraphics[width=0.95\textwidth]{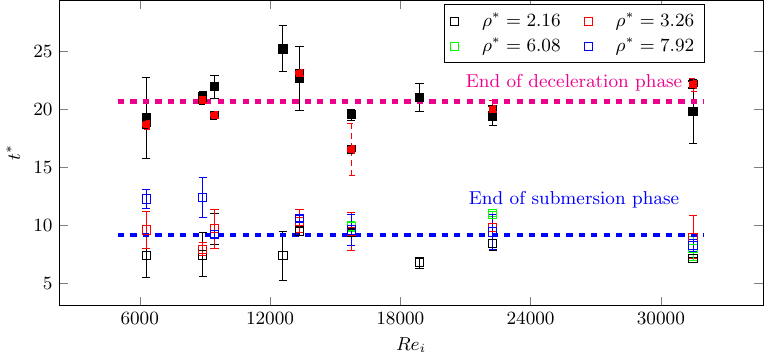}
    \caption{ (color online) Dimensionless time denoting end of submersion phase and end of deceleration phase of sphere motion as a function of initial Reynolds number and density ratio. 
     The error bars express one standard deviation computed from four repetitions of the same experiment. For densities $\rho^* = 6.08$ and $7.92$ of the 10~mm sphere, the termination of the deceleration phase is not captured, due to the sphere moving beyond the field of view. For the density ratio 7.92, the termination of the deceleration phase for the 4~mm and 6~mm sphere was also not clear, since the velocity did not exhibit a distinct minimum. }
    \label{fig:submersion phase boundary}
\end{figure}

The dimensionless time boundaries between the three phases of motion -- submersion, deceleration, settling -- are remarkably constant over initial Reynolds number, sphere diameter, and density ratio, as shown in Fig.~\ref{fig:submersion phase boundary} for the dimensionless time at which the submersion phase and deceleration phase ends. \citet{truscott2012unsteady} also measured the penetration trajectory of hydrophilic spheres with various density ratios,  all impacting with the same Reynolds number. Their measured dimensionless lateral displacement is in excellent agreement with the end of the submersion phase shown in Fig.~\ref{fig:submersion phase boundary}. Only one sphere trajectory in the work of \citet{truscott2012unsteady} reached a clear termination of the deceleration phase (minimum velocity in their Fig. 6(b)), and this occurred at a dimensionless time of $t^*=18$, also in good agreement with our value.

\begin{figure}
    \begin{subfigure}[b]{0.45\textwidth}
        \centering
        \includegraphics[width=0.9\textwidth]{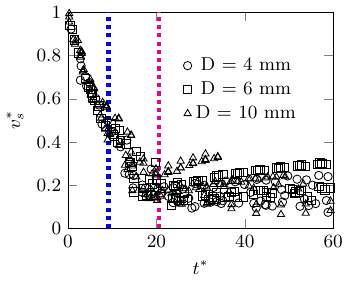}
        \caption{}
    \end{subfigure}
    \begin{subfigure}[b]{0.45\textwidth}
        \centering
        \includegraphics[width=0.95\textwidth]{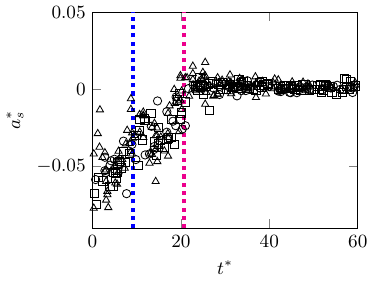}
        \caption{}
    \end{subfigure}
    \caption{Magnitude of dimensionless (a)  velocity ($v_s^*$) and (b) acceleration ($a_s^*$) for various sphere diameters and impact $Re_i$ ($\rho^* = 2.16$). The dashed blue line represents the end of the submersion phase, while the magenta dashed line illustrates the end of the deceleration phase.}
    \label{fig:kinematics_all}
\end{figure}

The similarity among the many experiments is further underlined in Fig.~\ref{fig:kinematics_all}, where the dimensionless magnitude of velocity and acceleration of spheres with a density ratio of $\rho^* = 2.16$ over time and for all sphere diameters and initial Reynolds numbers are shown.  Although the previous results in Figs.~\ref{fig:sample repetetion}(a)-(c) and \ref{fig:sample density deviations} indicate large variations in specific trajectories, in dimensionless terms, both the velocity and acceleration exhibit remarkable uniformity, confirming that the dimensionless scaling is appropriate. Furthermore, this uniformity is evident also for the other sphere density ratios. As such, these universal curves represent a predictive tool for estimating the deceleration of penetrating spheres within the parameter limits described above.

\subsection{Dynamics of sphere motion}
\label{subsection:dynamics}
\begin{figure}

   \begin{subfigure}[b]{0.35\textwidth}
    \includegraphics[width=\textwidth]{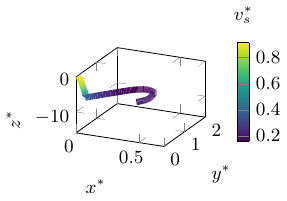}
 \caption{}
   \end{subfigure}
   \begin{subfigure}[b]{0.3\textwidth}
   \includegraphics[width=0.9\textwidth]{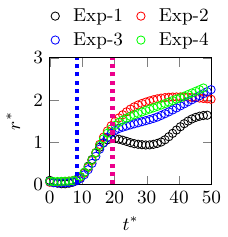}
 \caption{}
   \end{subfigure}
     \begin{subfigure}[b]{0.32\textwidth}
 \includegraphics[trim=0.1cm 0 0 0cm, width=0.9\textwidth]{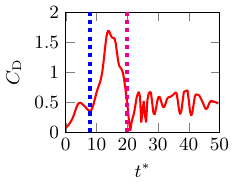}
         \caption{}
    \end{subfigure}
    
    \vspace{2em} 
    
    \begin{subfigure}{0.32\textwidth}
 \includegraphics[width=0.9\textwidth]{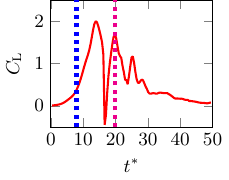}
         \caption{}
    \end{subfigure}
    \begin{subfigure}{0.33\textwidth}
 \includegraphics[width=0.95\textwidth]{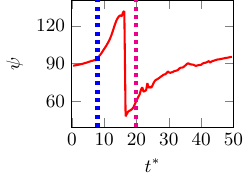}
         \caption{}
    \end{subfigure}
 \begin{subfigure}{0.32\textwidth}
 \includegraphics[width=0.9\textwidth]{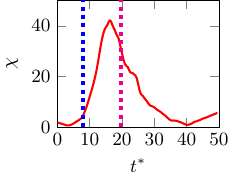}
         \caption{}
    \end{subfigure}
    \caption{ (color online) Penetration of a 10~mm sphere of density ratio $\rho^*=2.16$ at $Re_i=22300$. (a) Three-dimensional rendition of pathline in dimensionless time; (b) Dimensionless lateral distance $r^*$ for four repetitions of the same experiment; (c) Drag coefficient as a function of dimensionless time. (d) Lift coefficient as a function of dimensionless time. (e) and (f) represent angle $\psi$ and angle $\chi$ as a function of dimensionless time, respectively. The graphs (c)-(f) correspond to the red curve shown in graph (b). The color bar depicted in (a) illustrates the variation in dimensionless velocity (${v_s}^*$) of the sphere. The time step between the two successive points plotted in (b) is 6~ms. }
    \label{fig:Result 10 mm Re 22257}
\end{figure}

In discussing the dynamics of sphere motion, we again begin by examining an example set of data, in this case pertaining to the 10~mm sphere impacting at a Reynolds number of 22300, and shown in Fig.~\ref{fig:Result 10 mm Re 22257}. In this figure and in subsequent figures, for clarity, the graphs are plotted only with lines and no symbols. For each of the experimental conditions, four repetitions were performed, and the entire digitized trajectories ($x$, $y$, $z$ coordinates) over time are available and documented in \cite{billa_motion_2024}. Although there exists a high degree of randomness in the actual trajectories for each experimental repetition, the strong commonalities alluded to above will now be discussed in terms of the three phases, beginning with the submersion phase.\\

\textit{Submersion phase}

Examining Fig.~\ref{fig:Result 10 mm Re 22257}, in the submersion phase, the drag coefficient begins at zero, corresponding to the first contact of the lower sphere surface with the pool free surface. As the sphere submerges into the pool, the drag coefficient increases over the time it takes the sphere to submerge, approximately 4-5 diameters. 
At this time, the drag coefficient reaches a value typical for steady-state flow, i.e., $C_\mathrm{D} \approx 0.5$, and retains approximately this value until the end of the submersion phase. Interestingly, and referring back to Fig.~\ref{fig:submersion phase boundary}, this value of $t^*\approx$ 4-5 is virtually constant for all density ratios and Reynolds numbers, whereby Reynolds number has been varied through both impact velocity and sphere diameter. It appears, therefore, that upon entry, and independent of all impact parameters, the boundary layer on the sphere requires a translation of about 4-5 diameters to become fully developed to a stage devoid of water entry effects.

The distinguishing feature of the submersion phase is that the trajectory remains nearly vertical, meaning that during this phase, there are only very weak lift forces acting. This is quantitatively confirmed in Fig.~\ref{fig:Result 10 mm Re 22257}(d). According to the interpretation given above, this means that the boundary-layer separation and wake remain axisymmetric over this period of penetration. Any instability in the boundary layer or unsteadiness in the wake that would break this symmetry requires at least this dimensionless time to develop and become influential to the trajectory.  Corresponding to the vertical trajectory downward, the angles $\psi$ and $\chi$ both exhibit values which remain approximately constant at respectively 90$^\circ$ and 0$^\circ$ throughout this phase.\\

\textit{Deceleration phase}

We now move to the deceleration phase, marked by the first deviations from the initial vertical trajectory (see Figs.~\ref{fig:Result 10 mm Re 22257}(a) and ~\ref{fig:Result 10 mm Re 22257}(b)). The trajectory begins this phase with an abrupt curvature, which is reflected in the rapidly changing angles $\psi$ (Fig.~\ref{fig:Result 10 mm Re 22257}(e)) and $\chi$ (Fig.~\ref{fig:Result 10 mm Re 22257}(f)), evidently a result of a rapid increase in both drag and lift. Since the first trajectory curvature arises from the vertical state, the lift at this stage can only be attributed to hydrodynamic lift, since the body forces are acting vertical. Assuming the wall shear stress contributes little to the overall drag or lift, the integrated pressure over the sphere with a skewed wake would result in a drag and lift that would no longer be aligned with the motion axis of the sphere. The large values of drag and lift in this phase are, therefore, clearly related to asymmetric wake effects since these would result in a skewed base pressure area on the rear of the sphere.

Whereas the drag coefficient rises to a maximum value and then decreases again towards the end of the deceleration phase (Fig.~\ref{fig:Result 10 mm Re 22257}(c)), the lift coefficient exhibits an abrupt jump from positive to negative values (Fig.~\ref{fig:Result 10 mm Re 22257}(d)). This is very typical of all other data sets and can be explained by examining changes in the angle $\psi$. As discussed in the appendix, when the trajectory goes through a 'projected' inflection point, this angle can exhibit sharp jumps in magnitude, since the orientation of curvature, i.e., $\mathbf{n}_\sigma$, will change direction. Such a change is seen in Fig.~\ref{fig:Result 10 mm Re 22257}(e), corresponding to a sharp drop in lift coefficient. We conclude that this arises due to a reorientation of the wake, such that the asymmetry changes orientation on the sphere. During this change, the wake is momentarily symmetric, leading to a short period of zero hydrodynamic lift. However, whereas the wake \textit{orientation} is changing, the wake \textit{area} apparently remains larger, resulting in a persistently large drag coefficient, i.e.,  the wake base pressure still acts over an area undiminished in magnitude. Once through the point of changing the sign of $\mathbf{n}_\sigma$, the lift force is again high and in the direction of the new unit vector -$\mathbf{n}_\sigma$, i.e., positive.

What is particularly noteworthy is that the lateral deviation $r^*$ exhibits a constant slope over the entire deceleration phase, almost identical in all repetitions of experiments at the same initial conditions. This infers a sustained hydrodynamic lift force in a constant direction. This deduction must be explained. The net gravity and buoyancy force acts downward, and would act to make the trajectory vertical. If the sphere is not moving vertically downward, then these body forces contribute to lift through angle $\psi$. Whereas a vertical trajectory results in a constant value of $r^*$,  this is not observed in the deceleration phase. Therefore, over this entire deceleration phase, there must be a sustained and constant hydrodynamic lift force in both magnitude and direction, counteracting the ever-present body force acting downward. This suggests that the asymmetric vortex shedding, which is understood to be the origin of the lift force, once established, remains approximately constant in its orientation on the sphere throughout this deceleration phase.

\begin{figure}
    \begin{subfigure}[b]{0.34\textwidth}
    \includegraphics[width=0.9\textwidth]{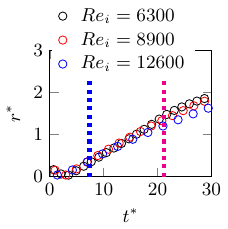}
 \caption{}
   \end{subfigure}
   \begin{subfigure}[b]{0.3\textwidth}
   \includegraphics[width=0.9\textwidth]{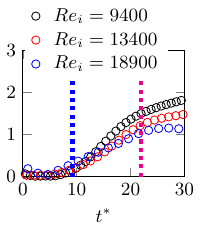}
 \caption{}
   \end{subfigure}
     \begin{subfigure}[b]{0.3\textwidth}
 \includegraphics[width=0.9\textwidth]{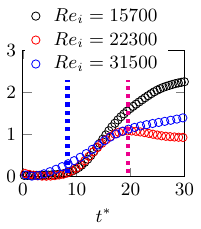}
         \caption{}
    \end{subfigure}
    
    \vspace{2em} 
    
    \begin{subfigure}[b]{0.32\textwidth}
 \includegraphics[trim=0.3cm 0 0 0cm, width=\textwidth]{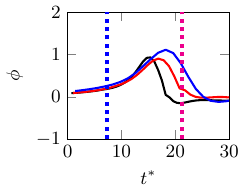}
         \caption{}
    \end{subfigure}
    \begin{subfigure}[b]{0.3\textwidth}
 \includegraphics[width=\textwidth]{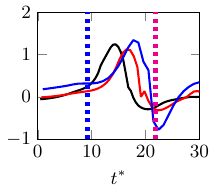}
         \caption{}
    \end{subfigure}
 \begin{subfigure}[b]{0.3\textwidth}
 \includegraphics[width=\textwidth]{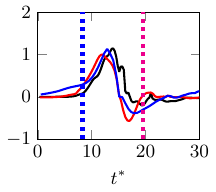}
         \caption{}
    \end{subfigure}
    \caption{(color online) Lateral deviations ($r^*$) and the ratio of the vertical component of the hydrodynamic lift force to the total body forces ($\phi$) are shown for various initial Reynolds numbers. Results are shown for spheres of diameter $D =4 $ mm in (a, d), $D = 6$ mm in (b, e), and $D = 10$ mm in (c, f). In (a, b, c), the time interval between successive points is 4 ms. For clarity, $\phi$ is plotted as continuous lines in (d, e, f). }
    \label{fig:Lift_Body_Separate_2.16}
\end{figure}

In this regard, the behaviour of the three different sized spheres shown in Fig.~\ref{fig:sample repetetion} at the end of the deceleration phase and in the settling phase is distinctly different. Whereas the trajectories of the 10~mm sphere (see Fig.~\ref{fig:sample repetetion}(c)) tend toward a vertical motion ($r^*$ constant) following the deceleration phase, the 6~mm sphere (see Fig.~\ref{fig:sample repetetion}(b)) exhibits larger fluctuations in trajectory inclination, albeit fluctuating around an approximate constant value. The 4~mm sphere (see Fig.~\ref{fig:sample repetetion}(a)) continues on an inclined trajectory throughout the entire field of view. In the decelerating phase, the body forces are competing with the hydrodynamic lift force to achieve a vertical trajectory, as described above. Having computed the hydrodynamic lift force coefficient at all times, the ratio of the vertical component of the hydrodynamic lift force to the sum of the body forces can be given as
\begin{equation}
    \phi=\frac{F_\mathrm{HL}\cos{(180-\psi)}}{F_\mathrm{G}-F_\mathrm{B}}
    = \frac{3}{4g}\frac{C_\mathrm{L} v_\mathrm{s}^2 \cos{(180-\psi)}}{D(\rho^* -1)}
    \label{eq:phi}
\end{equation}
 and is shown for the three sphere diameters 4, 6, and 10~mm, and for varying initial Reynolds numbers in Fig.~\ref{fig:Lift_Body_Separate_2.16}. From this diagram, this value does tend toward unity throughout the deceleration phase. From Eq.~(\ref{eq:phi}) it can be seen that the body forces will tend to dominate for larger spheres and from the diagram, the values for the 10~mm sphere appear to decrease earlier in dimensionless time, suggesting that the body forces begin to dominate the hydrodynamic lift forces earlier for the heavier sphere. This is intuitively correct, since the sphere is decelerating and the hydrodynamic lift force scales with velocity squared and diameter squared, whereas the body forces scale with diameter cubed and remain constant. What remains unclear is the exact reason that such a stable hydrodynamic lift force, i.e., an asymmetric wake, is maintained throughout the deceleration phase, and for the 4~mm sphere even throughout the settling phase. Further insight would require a time resolved measurement of the velocity field around the sphere throughout this phase.

Phenomenologically, these observations are in good agreement with experiments from \citet{truscott2012unsteady}, where also all hydrophilic spheres initially exhibited a sharp deviation from the vertical, accompanied by rapid deceleration. Similarly, they attribute the trajectory change from the vertical to asymmetrical vortex shedding, as observed and explained also by \citet{horowitz2010effect}. The exact direction of trajectory change was non-repeatable, as in the present experiments. What the present experiments reveal, is that while the direction of trajectory change is non-repeatable, the growth of lateral displacement remains constant in dimensionless time until a minimum velocity is reached, as indicated in Fig.~\ref{fig:kinematics_all}(a).

The deceleration phase ends when the absolute velocity magnitude reaches a minimum, and again, this occurs for all investigated cases at an approximately constant value of $t^* \approx$ 20 - 21, as seen in Fig.~\ref{fig:submersion phase boundary}. Typically, this minimum velocity is close to and sometimes slightly less than the terminal velocity, given in Eq.~(\ref{eq; terminal velocity}). For instance, in Fig.~\ref{fig:sample kinematics} for the 10~mm sphere, a slight acceleration is seen following the deceleration phase. This is likely due to a relaxation of the asymmetric wake into a symmetric, steady state condition, which is then maintained throughout the final settling phase.   Note that \citet{truscott2012unsteady}  also observe for lighter spheres a velocity slightly less than the terminal velocity, followed by a light acceleration (their Fig. 5). They call this `underdamped behaviour', but do not elaborate on its physical origins.

 Assuming the end of the deceleration phase corresponds to conditions at which inertia no longer contributes significantly to the force balance (Eq.~\ref{eq:Force evaluation}), it can be concluded that the time to dissipate the initial kinetic energy scales remarkably well with $D/U$,  independent of initial Reynolds number and density ratio. Retrospectively, this is not surprising, since no other length or velocities scales are involved in the problem.

\begin{figure}
\centering
    \begin{subfigure}[b]{0.32\textwidth}
        \centering
        \includegraphics[width=0.9\textwidth]{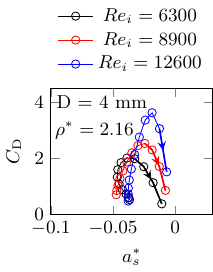}
 \caption{}
   \end{subfigure}
   \hfill
 \begin{subfigure}[b]{0.32\textwidth}
        \centering
        \includegraphics[width=0.9\textwidth]{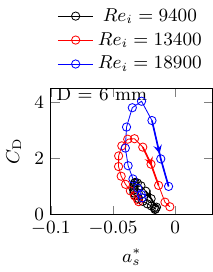}
 \caption{}
   \end{subfigure}
    \hfill
   \begin{subfigure}[b]{0.32\textwidth}
        \centering
        \includegraphics[width=0.9\textwidth]{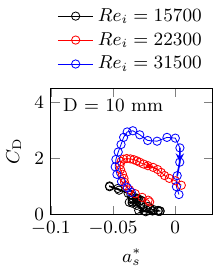}
 \caption{}
   \end{subfigure}
    
    \vspace{0.5cm}
    
   \begin{subfigure}[b]{0.32\textwidth}
        \centering
        \includegraphics[width=0.9\textwidth]{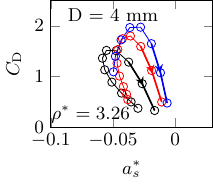}
 \caption{}
   \end{subfigure}
   \hfill
 \begin{subfigure}[b]{0.32\textwidth}
        \centering
        \includegraphics[width=0.9\textwidth]{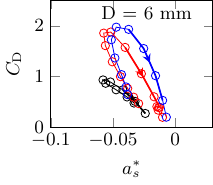}
 \caption{}
   \end{subfigure}
    \hfill
    \begin{subfigure}[b]{0.32\textwidth}
        \centering
        \includegraphics[width=0.9\textwidth]{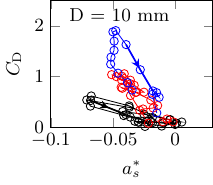}
 \caption{}
   \end{subfigure}
    \caption{(color online) Dependence of instantaneous drag coefficient on instantaneous deceleration for times within the deceleration phase. (a,b,c) $\rho^*=2.16$; (d,e,f) $\rho^*=3.26$. The arrows indicate an increasing time in the  approx. interval $t^*=10 - 20$. The legends in (a), (b), and (c) apply to (d), (e), and (f), respectively. }
    \label{fig:C_D vs a with lines}
\end{figure}

It is now consequential to plot the drag coefficient evolution as a function of deceleration. This is done for the two spheres $\rho^*=2.16$ and 3.26  in Fig.~\ref{fig:C_D vs a with lines} for all instantaneous values within the deceleration phase, i.e., excluding the submersion and settling phases, assuming these phases with residual entry effects or with no inertial effects may exhibit a different behaviour. Although some irregularity and scatter are observed in this data, all graphs show a very similar pattern: first, increasing $C_\mathrm{D}$ with increasing deceleration followed by decreasing $C_\mathrm{D}$ with decreasing deceleration. The following is postulated from this data. Under initial deceleration at the beginning of the deceleration phase, the drag coefficient increases due to the earlier boundary-layer separation, as described above, thus increasing the deceleration. This is then a self-augmenting and perpetuating process, which continues until some limit is reached, close to the maximum drag coefficient. Thereafter, any perturbation causing the boundary layer to attach later will decrease the wake area, the drag coefficient and the deceleration; again a self-augmenting process, until the sphere reaches its minimum velocity at the end of the deceleration phase. As seen in this figure, the limiting or maximum drag coefficient is larger for the lighter sphere ($\rho^*=2.16$), presumably since inertia is more dominant for the heavier sphere ($\rho^*=3.26$). This then explains the rise and fall of the drag coefficient seen in the deceleration phase, for instance, shown exemplary in Fig.~\ref{fig:Result 10 mm Re 22257}. For the spheres, $\rho^*=6.08$ and 7.92 such curves are not obtained simply because the inertial force overwhelms any small variation of deceleration.

This self-augmenting behaviour is inherent to any instability and gives rise to hysteresis. This is evident in the curves of Fig.~\ref{fig:C_D vs a with lines}, where for the same value of deceleration, two values of the drag coefficient can be obtained, depending on whether the deceleration is increasing or decreasing. This is, therefore, a history effect arising from the boundary-layer separation adjusting to varying values of the sphere velocity; however, this history effect is more influential than the Boussinesq-Basset term, which only considers viscous drag and not pressure drag, as in this case.

\begin{figure}
    \centering
    \includegraphics[width=0.9\textwidth]{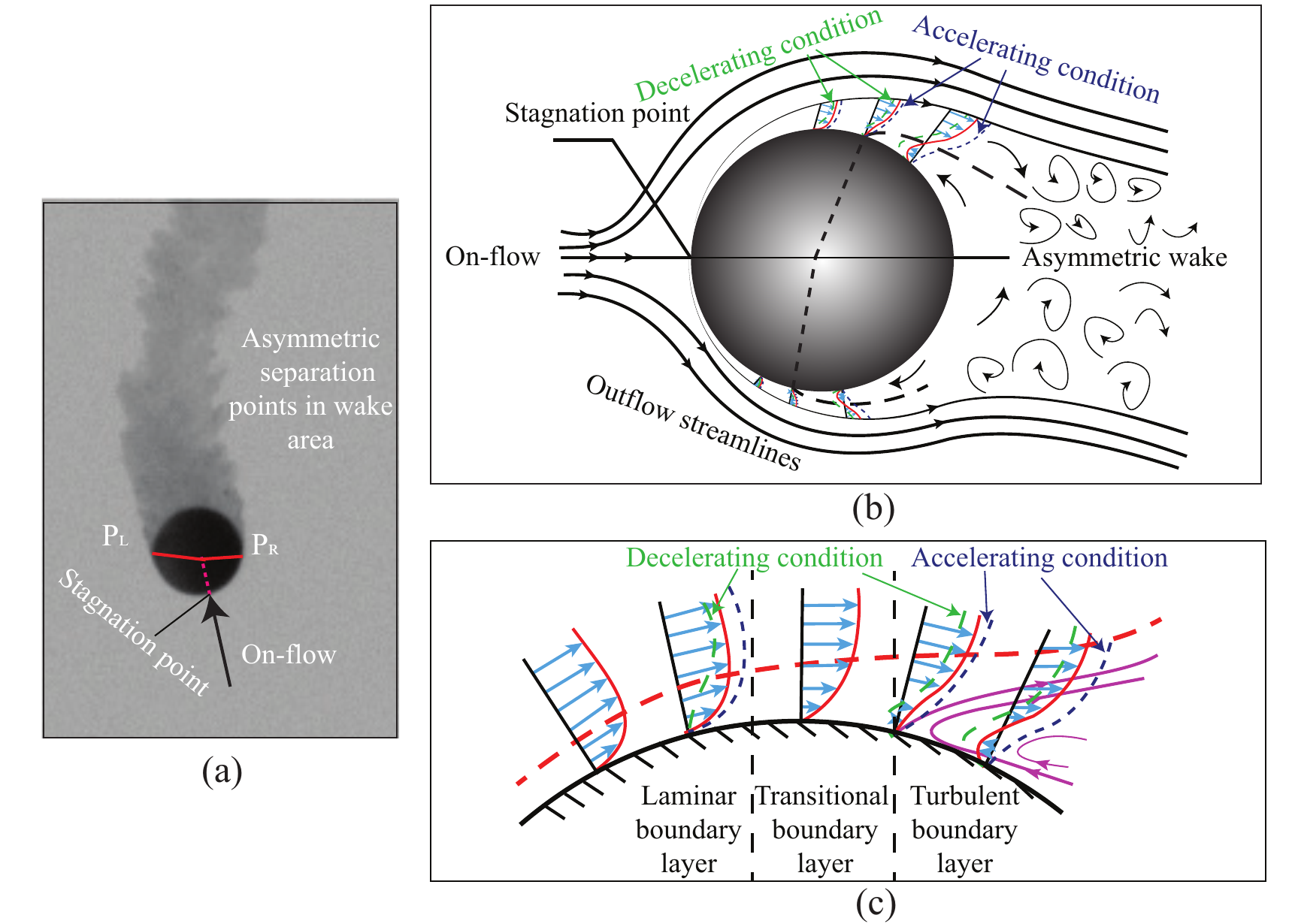}
    \caption{(color online) Schematic interpretation of the origins of lift and drag through wake asymmetry. (a) Visualization of a free-falling sphere ($\rho^* = 2.16$, $D = 10 $ mm and $Re_i = 15700$), showing a change in trajectory and indicating an asymmetry of the left (P$_\mathrm{L}$) and right (P$_\mathrm{R}$) points of separation. (b) Schematic (and exaggerated) visualization of asymmetric wake area arising from asymmetric separation points, indicating that relative to the flow direction, a lift force component arises. (c) Schematic (and exaggerated) interpretation of how a deceleration or acceleration of the sphere might influence the boundary-layer velocity profile in the vicinity of a separation point. }
    \label{fig:BL}
\end{figure}

To explain the above results, we now consider the flow around the sphere under conditions of strong acceleration/deceleration, particularly the behaviour of the boundary layer near its separation point. To observe the flow patterns around the sphere, a uniform coating of red dye was applied to the sphere before it was released into the quiescent liquid with a $Re_i$ of 15700. Figure.~\ref{fig:BL}(a) shows the resulting flow visualization around the sphere.  Under steady flow conditions (which is not the case in the deceleration phase), separation for a laminar boundary layer would be expected near an angle of 80$^\circ$-84$^\circ$.
The expected velocity profile of the boundary layer in the vicinity of separation is shown schematically in Fig.~\ref{fig:BL}(c), whereby the solid lines represent the boundary layer velocity profile at some instant in time. If the sphere experiences a strong acceleration, the outer flow would effectively increase in velocity, and over a short period of time, the separation point would move rearward, decreasing the wake area. This is indicated in the figure with a dashed line marked `accelerating sphere'. With a deceleration of the sphere, the opposite trend would be expected, i.e., the separation would occur earlier, resulting in a larger wake area over which the base pressure would be exerted. In the figure, this is indicated with the dashed velocity profile labelled `decelerating sphere'. This would result in a larger base pressure area; hence, a higher drag, consistent with the observations of the present study throughout the deceleration phase. \\
It is apparent from the data in Fig.~\ref{fig:C_D vs a with lines}, that the drag coefficient of a decelerating sphere at some instantaneous Reynolds number can be significantly higher than for a sphere experiencing a steady flow at the same  Reynolds number, usually of the order 0.5. Although there exist numerous studies of sphere drag under decelerating conditions, most of these are devoted to low Reynolds number flows \citep{liu2018numerical,velazquez2024simplified,temkin1982droplet}. Some studies of decelerating bluff bodies at similar Reynolds numbers have indicated that the drag coefficient can differ significantly from values for steady flow \citep{potvin2003new}, while others remain rather inconclusive about how drag changes for decelerating bodies \citep{marchildon1979effects}. Thus, at present it is difficult to find corroborating data to our experimental results.

Similarly, little quantitative literature exists on unsteady lift forces in this Reynolds number range, as confirmed in the exhaustive survey of spherical particles in unbounded flows given by \cite{shi2019lift}. Nevertheless, it is clear that bifurcations and symmetry breaking in the wake of spheres occur already at relatively low Reynolds numbers and result in a random change of direction \citep{fabre2008bifurcations}. The magnitude of these lift forces in the present Reynolds number range has, to the authors' knowledge, not been previously measured. 

This transient state of the boundary layer during sphere deceleration is analogous to Stokes first problem in which the similarity variable $\eta=y/2\sqrt{\nu t}$ is used to describe the invoked velocity profile above a plate, suddenly accelerated to some finite velocity  \citep{spurk2007fluid}. Interpreting $y$ as the boundary-layer thickness in the present case, the time necessary for the diffusion of a strong change of outer flow velocity into the same dimensionless state near the sphere surface would scale with $y^2$. If, for instance, one boundary layer was only half as thick as a second boundary layer, a change of outer velocity would be felt in $0.5^2$, i.e., a quarter of the time. Given that the boundary-layer thickness over the sphere will scale approximately with Re$^{1/2}$, the spheres with higher initial Reynolds numbers would exhibit a lower boundary-layer thickness. Hence, this should lead to an earlier and stronger influence on the separation point, resulting in a higher drag. This is consistently seen in the data of Fig.~\ref{fig:C_D vs a with lines}.

The spheres with a higher density ratio will have higher inertial forces for the same Reynolds number; hence, any wake variations will have a relatively lower effect on the overall sphere motion, resulting in less modulation of the computed drag coefficient using Eq.~(\ref{eq:Force evaluation}). The results shown in Fig.~\ref{fig:C_D vs a with lines} confirm that spheres with higher density ratios exhibit lower maximum drag coefficients.

\textit{Settling phase}
The settling phase is prominently characterised with widely varying values of lateral deviation $r^*$, as shown in Fig.~\ref{fig:sample repetetion}, but also evident in Fig.~\ref{fig:Result 10 mm Re 22257}(b). During this phase, the velocity is very low, and inertia no longer plays a dominant role. Thus, the trajectory is now highly susceptible to even small lift forces arising from wake asymmetries. Thus, the trajectory curvatures become more frequent and pronounced. This leads to variations in drag coefficient, but now only seen as smaller variations near values typical of steady-state flow, i.e., $C_\mathrm{D} \approx 0.5$.  Similarly, the lift coefficient reduces in magnitude, approaching values near zero. These variations, both in drag and lift, are still attributed to a fluctuating and non-symmetric wake structure, but its consequence is now much less significant.

Referring back to the fluid mechanic interpretation of the deceleration phase and in particular to the yet unexplained stability of the wake asymmetry orientation throughout this phase, it is evident that this stability no longer exists in the settling phase. Not only the dimensionless lateral displacement exhibits large fluctuations, but also the orientation angles $\psi$ and $\chi$ are continually changing. This is seen in Fig.~\ref{fig:Result 10 mm Re 22257}, but is also evident in Fig.~\ref{fig:sample repetetion}(d)-(i), in which similar results are shown for several other spheres and impact conditions. Recalling the physical meaning of the angles $\psi$ and $\chi$ given above in section~\ref{subsection:kinematics}, it is evident from these measurements that in this phase, after which the velocity has reached a minimum value, the vortex shedding and instantaneous wake and associated base pressure area exhibit much less stability in their orientation w.r.t. the flow direction. The continual variations of $\psi$ and $\chi$, while small in magnitude, suggest a random change in pathline orientation, i.e., a random change in lift force orientation. This would appear to be a consequence of random asymmetry directions of the wake and/or vortex shedding. Given the low velocities and correspondingly low inertial forces, even small changes in wake orientation will have a larger effect on the trajectory in this settling phase.

\section{Conclusions}
\label{sec:conclusions}
In this study we have introduced the analysis of free body trajectories in a natural coordinate system, allowing instantaneous lift and drag forces to be measured independent of one another. This technique has been applied to study spheres penetrating a liquid pool. We have associated measured changes in drag with changes of the base pressure wake area and changes of lift with variations of the wake and base pressure area orientation with respect to the current flow direction. We attribute such changes of wake area and orientation to vortex shedding. The vortex shedding remains highly axisymmetric over a uniform dimensionless submersion time, independent of initial Reynolds number or sphere-to-liquid density ratio. In a subsequent deceleration phase the vortex shedding exhibits a constant, but skewed orientation to the flow direction, counteracting the downward acting body forces. This observation is a consequence of the  inferred constant hydrodynamic lift force in this phase. In a final settling phase, in which inertial forces are no longer significant, the vortex shedding becomes more random in orientation, both for single trajectories and over repeated experiments, leading to irregular, albeit low variations in lift and drag.

A further result relates to the very high drag coefficients measured during the deceleration phase of the spheres. Having associated drag with base pressure area to a first approximation, an explanation for this behaviour is postulated on the basis of boundary-layer separation. Specifically, the boundary layer is assumed to separate on average earlier from the sphere under decelerating conditions, and later for accelerating conditions. The former would lead to a larger base pressure area; hence, a higher drag, as measured. The examination of drag coefficient as a function of deceleration reveals a distinct self-augmenting behaviour, suggesting an instability and leading to hysteresis of drag coefficient with deceleration gradient. Associated with the higher drag under decelerating conditions are high lift forces, albeit not with a constant orientation. This suggests that the boundary-layer separation occurs not only earlier under decelerating conditions but also no longer completely axisymmetric. This occurs especially over the period of the deceleration phase, again emphasizing that this phase duration in dimensionless time is approximately the same for all initial conditions and sphere density ratios.

Of practical use is the fact that all sphere experiments follow virtually identical velocity and acceleration traces when plotted in dimensionless time. This offers a predictive tool in knowing how far the sphere will penetrate in what time and what forces will be exerted on it during this period. 
Finally, the study provides an openly available and comprehensive data set of all experiments \citep{billa_motion_2024}.

This work postulates very specific behaviour of the time dependent flow separation from the decelerating sphere, derived from the measured forces. In principle, this flow separation could be quantified using a whole field measurement technique such as time resolved particle tracking velocimetry, although this would require a high seed particle density, high spatial resolution, a large field of view, and appropriate processing, e.g., shake-the-box. Nevertheless, the present quantitative observations serve as motivation for subsequent validation through such a measurement.

\section*{Appendix}
\begin{figure}
    \centering
    \includegraphics[width=0.9\textwidth]{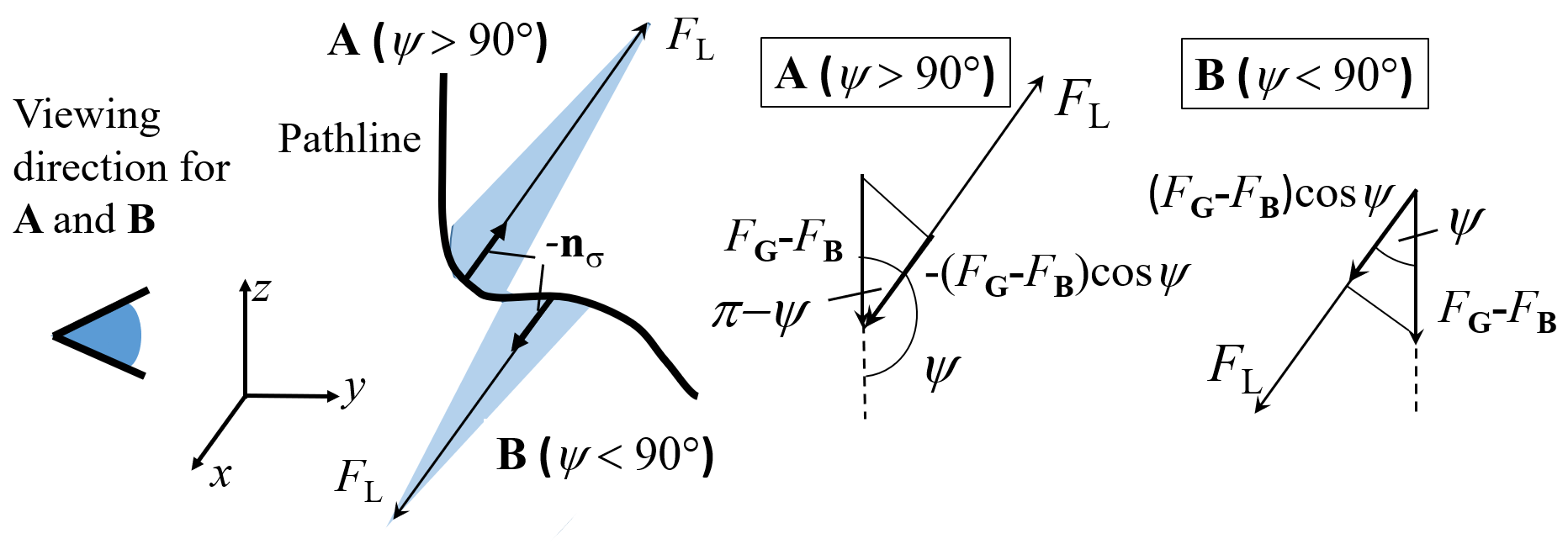}
    \caption{(color online) Graphical representation of example lift force decomposition for two curvatures \textbf{A} ($\psi>\pi/2$) and \textbf{B} ($\psi<\pi/2$) on opposite sides of an inflection point of the sphere pathline.  }
    \label{fig:inflection point}
\end{figure}

In anticipation of properly interpreting the measurement results, the decomposition of the total lift force into body forces and a hydrodynamic lift force is graphically represented in Fig.~\ref{fig:inflection point}. A path line is shown, which begins vertically downward, exhibits a curvature with $\psi>\pi/2$, and then changes curvature orientation. If this curvature change constitutes an inflection point when projected onto any vertical plane, then this leads to a jump of the $\mathbf{n}_\sigma$ vector by up to -$\pi$, for example, in Fig.~\ref{fig:inflection point} to $\psi<\pi/2$. This also leads to abrupt jumps in lift coefficient, e.g. Fig.~\ref{fig:Result 10 mm Re 22257}(d). This is not a physical jump in lift in magnitude, only a consequence of the coordinate definitions.

\section*{Acknowledgement}
    The author, C. Tropea, acknowledges support from the Indian Institute of Technology Madras through the appointment as the Henry Ford Chair Professor in Mechanical Engineering. The authors would like to thank Prof. Arvind Pattamatta for providing access to the high speed camera.

\section*{Declaration of interests}
The authors report no conflict of interest.

\bibliographystyle{jfm}
\bibliography{references_new}

\end{document}